\documentclass[12pt,a4paper]{article}
    \usepackage{graphicx,amsmath,amsfonts}

    \makeatletter
    \@addtoreset{equation}{section}
    \makeatother
    \renewcommand{\theequation}{\arabic{section}.\arabic{equation}}

\begin{document}
\title{Semiclassical gravitational effects\\ near a singular magnetic flux}
\author{Volodymyr M. Gorkavenko\thanks{E-mail:
gorka@univ.kiev.ua},\\ \it \small Department of Physics, Taras
Shevchenko National University of Kyiv,\\ \it \small 6 Academician
Glushkov ave., Kyiv 03680, Ukraine
\\ \phantom{11111111111}\\
Alexander V. Viznyuk\thanks{E-mail: sv@mail.univ.kiev.ua}\\
\it \small Bogolyubov Institute for Theoretical Physics,
\\
\it \small 14-b Metrologichna str., Kyiv 03143, Ukraine}
\date{}
\maketitle

\abstract{We consider the backreaction  of the vacuum polarization
effect for a massive charged scalar field in the presence of a
singular magnetic massless string on the background metric. Using
semiclassical approach, we find the first-order (in $\hbar$ units)
metric modifications and the corresponding gravitational potential
and deficit angle. It is shown that, in certain region of values
of coupling constant and magnetic flux, the gravitational
potential and deficit angle can be positive as well as negative
over all distances from the string and can even change its sign.
Unlike the case of  massless scalar field, the gravitational
corrections were found to have short-range behavior.}

\section{Introduction}
Gauge theories with spontaneous symmetry breaking  predict the
emergence of cosmic objects with topology defect in the early
Universe. Such objects can possibly survive at the present day
(see the review by Vilenkin\cite{AV1} and references therein). In
topology defect points the spontaneous symmetry breaking
principle, giving the mass for fields, is no more valid. So
physical fields need some boundary conditions, that cause the
vacuum polarization and appearance of non-zero vacuum expectation
value of the energy momentum tensor of quantum fields like in
Casimir effect \cite{Cas}. Non-zero vacuum expectation value of
the energy momentum tensor in one's turn serves as  a source of
gravitation \cite{Bir,Most} and can take part in cosmological
models of the Universe taking into account vacuum quantum effects.

One of the topology defect manifestations, whose existence is not
in contradiction with observable data\cite{Kibble}, is cosmic
strings which are particularly interested both as possible "seed"
for galaxy formation \cite{AV2,KibTur} and as possible
gravitational lens \cite{AV3}. Space-time metric of the cosmic
strings in empty Universe in the linearized approach were found by
Vilenkin \cite{AV3} and exactly in \cite{Hich,Many}. In spite of
large linear mass density of the string $\mu$ ($\sim 10^{22}$
g/cm) the space-time metric is not highly curved near the string
and for a static, cylindrically symmetric cosmic string is conical
and hence flat\footnote{Here and over all the paper we use  Plank
units: $G=\hbar=c=1$ in which $\mu\sim10^{-6}$.}
\begin{equation}\label{metr}
ds^2=-dt^2+dz^2+dr^2+(1-4\mu)^2r^2d\varphi^2
\end{equation}
with deficit angle $\triangle\varphi=8\pi\mu$. The effects of the
quantum-mechanical scattering of a test particle on a string were
estimated in \cite{Hooft}-\cite{Misch}. The vacuum polarization
effect of quantum fields in the string background is considerably
large near the string (see for example \cite{Many1}). So it can
significantly modify space-time metric in the vicinity of the
string. The backreaction of the vacuum energy momentum tensor on
space-time metric was first investigated by Hiscock \cite{H} in
linear perturbations within the semiclassical approach. If the
cosmic string carries a magnetic flux, the vacuum polarization has
also contribution from the Bohm-Aharonov interactions
\cite{Aha,Sereb}. In this case, the vacuum expectation value of
the energy momentum tensor of quantum fields was derived both for
massless \cite{Many2} and  massive field \cite{SR,SG}, but
backreaction of the vacuum polarization was analyzed in detail
only for massless field \cite{G}.

As known, the energy momentum tensor in the case of material field
of zero mass is equivalent to the case of massive field at small
distances. But, as one can see from \cite{SG}, in the case of
massless material field the physical peculiarities of the tensor
components behavior at finite distances are lost. In particular,
the short-range exponentially decreasing of the tensor components
is absent, and, more interesting, the tensor components of the
massless field in principle lose possibility of changing its sign
under moving away from the string. Hence, the case of the massless
field is a first order approximation of the general massive case.
To illustrate above, we refer to \cite{SG}, where the massive
scalar field was detailed considered and the case of zero mass is
a simple constant on the figures for dimensionless tensor
component $(r^4t^{00}, r^4t^{rr}, r^6t^{\varphi\varphi},
r^4t^{33})$ instead of evidently complicated structure. In this
respect, it seems to be of interest to carefully consider the
backreaction in the common case of massive quantum field to see
possible new features arising from the massiveness.

In this paper, we use the analytically obtained result \cite{SG}
to explicitly investigate in the linear approximation  the
backreaction of the massive field on the space-time metric and
physical consequences of one.

In Section 2 we  generalize the  linear approximation method
\cite{H}  to the case of arbitrary field in the background of
cosmic string. Using results of Section 2 and \cite{SG}, in
Section 3 we analytically find expressions for the modified
metric, Newtonian gravitation potential and deficit angle, that
are analyzed in Sections 4, 5 in detail. Discussions of obtained
results can be found in Concluding remarks, some mathematical
aspects are placed in Appendix.


\section{Perturbative approach}
The exterior metric of a static, cylindrically symmetric cosmic
string (with or without the magnetic flux $\Phi$)  is
Eq.\eqref{metr}. This metric induces  non-zero vacuum expectation
values of the energy-momentum tensor $\langle
T^{\mu}_{\nu}\rangle$  of a quantum field. We are interested in
considering the backreaction of this energy-momentum tensor on a
string's metric. To do this in a semiclassical approach, one has
to solve the Einstein equations
 \begin{equation}\label{sma}
    G^{\mu}_{\nu}=8\pi\langle T^{\mu}_{\nu}\rangle.
\end{equation}

As we pointed out in Introduction, such problem for the massless
fields was first solved by Hiscock  in \cite{H}. In this Section,
we follow \cite{H} to obtain the result for any type of fields.

The general static, cylindrically symmetric and invariant under
Lorentz boosts along the z-axis metric has the form
\begin{equation}\label{gm}
    ds^2=e^{2\phi(r)}(-dt^2+dz^2+dr^2)+e^{2\psi(r)}d\varphi^2.
\end{equation}

Corresponding components of the Einstein tensor are
\begin{equation}\label{gt}
    G^{t}_{t}=G^{z}_{z}=e^{-2\phi}(\phi''+\psi''+(\psi')^2),
\end{equation}
\begin{equation}\label{gr}
    G^{r}_{r}=e^{-2\phi}((\phi')^2+2\phi'\psi'),
\end{equation}
\begin{equation}\label{gfi}
    G^{\varphi}_{\varphi}=e^{-2\phi}((\phi')^2+2\phi''),
\end{equation}
where prime means the derivative by $r$.

If $G^{\mu}_{\nu}=0$ in the exterior of a string, we have
\begin{equation}\label{0apr}
   \phi_0=0, \,\,\,\,\,\psi_0=\ln(\alpha r).
\end{equation}
To join interior and exterior solutions one have to put
$\alpha=1-4\mu$ (see \cite{Hich} for details) and we recover
\eqref{metr}.

Since $\langle T^{\mu}_{\nu}\rangle$ is small quantum correction,
we can expand the solution of \eqref{sma} about the background
metric \eqref{0apr}:
\begin{equation}\label{exp}
    \phi=\phi_0+\phi_1\,,\,\,\,\,\psi=\psi_0+\psi_1
\end{equation}
where $\psi_0$ and $\phi_0$ are from \eqref{0apr},  and $\phi_1$
and $\psi_1$ are supposed to be the first order of smallness, same
as $\langle T^{\mu}_{\nu}\rangle$.

 In the  first order approximation, Eq.\eqref{sma} takes form
 \begin{equation}\label{eqt}
    \phi_1''+\psi_1''+\frac{2}{r}\psi_1'=8\pi\langle
    T^{t}_{t}\rangle,
\end{equation}
\begin{equation}\label{eqr}
    \frac{2}{r}\phi_1'=8\pi\langle T^{r}_{r}\rangle,
\end{equation}
\begin{equation}\label{eqfi}
    2\phi_1''=8\pi\langle T^{\varphi}_{\varphi}\rangle,
\end{equation}
and the exterior metric \eqref{gm} modifies to
\begin{equation}\label{mla}
    ds^2=(1+2\phi_1(r))[-dt^2+dz^2+dr^2]+(1-4\mu)^2r^2(1+2\psi_1)d\varphi^2.
\end{equation}
Eqs.\eqref{eqr}-\eqref{eqfi} that define $\phi_1$ function are
adjusted if $\langle T^{r}_{r}\rangle+r(\langle
T^{r}_{r}\rangle)'=\langle T^{\varphi}_{\varphi}\rangle$ which is
just a r-component of the covariant conservation condition  for
the energy-momentum tensor ($\nabla_\mu\langle
T^{\mu}_{\nu}\rangle=0$). We propose it to be justified.

Using substitution $\psi_{1}'=\chi/r^2$, solution of
Eqs.\eqref{eqt}-\eqref{eqr} can be easily found:
\begin{equation}\label{phi1}
    \phi_1(r)=4\pi\int\limits^{r}_{\infty}dr'\cdot r'\langle T^{r}_{r}(r')\rangle,
\end{equation}
\begin{equation}\label{psi1}
    \psi_{1}(r)=4\pi\int\limits^{r}_{\infty}\frac{dr'}{r'^{2}}\int\limits^{r'}_{\infty}dr
    ''\cdot r''^{2}[2\langle T^{t}_{t}(r'')\rangle-\langle
    T^{\varphi}_{\varphi}(r'')\rangle].
\end{equation}
Lower limits of integration defined so the $\phi_{1}(r)$ and
$\psi_{1}(r)$ to be vanishing at infinity\footnote{We expect the
induced vacuum expectation values of energy-momentum tensor to be
decreasing function of distance from the string, so it is natural
to choose the constants of integration so the metric \eqref{mla}
is flat at infinity.}. In other words, we neglect the homogeneous
solution of \eqref{eqt}-\eqref{eqfi} as having no relation to our
effect.

It is more convenient to introduce new radial coordinate $\rho$,
which is the measure of proper distance from the string:
\begin{equation}\label{R}
    d\rho=\sqrt{1+2\phi_{1}(r)}dr \,\,\, \Rightarrow \,\,\, \rho\approx
    r+\int\limits^{r}_{\infty}\phi_{1}(r')dr',
\end{equation}
where we again chose the arbitrary constant so $\rho$ and $r$ to
be equal at infinity. With the same accuracy up to linear terms
\begin{equation}\label{r}
    r\approx \rho-\int\limits^{\rho}_{\infty}\phi_{1}(\rho')d\rho'\, ,
    \,\,\,\,\,\,\phi_{1}(r)\approx \phi_{1}(\rho)\,\,\,\,\,\, \rm{and} \,\,\,\,\,\,
    \psi_{1}(r)\approx\psi_{1}(\rho)
\end{equation}

Using \eqref{R} and \eqref{r} we can rewrite the induced metric
\eqref{mla} in the form
\begin{equation}\label{metric}
    ds^2=\left(1+2\phi_{1}(\rho)\right)[-dt^2+dz^2]+d\rho^2+(1-4\mu)^2\rho^2
    \left(1+2\psi_{1}(\rho)-\frac{2}{\rho}\int\limits^{\rho}_{\infty}
    \phi_{1}(\rho')d\rho'\right)d\varphi^2
\end{equation}
The condition of validity of this result is the smallness of first
order perturbation comparing to one:
\begin{equation}\label{cond of validity}
     |\phi_1(\rho)|\ll 1\,, \quad |\psi_1(\rho)|\ll1.
\end{equation}

Newtonian gravitational potential $V$ is recovered from the
$g_{00}$ component of the metric as $g_{00}=-(1+2V)$, so it is
$\phi_{1}(\rho)$ in our case. Gravitational force acting at the
probe particle with unit mass is
\begin{equation}\label{force}
    f(\rho)=-\phi_{1}'(\rho)=-4\pi \rho\langle T^{r}_{r}(\rho)\rangle
\end{equation}
where we used \eqref{eqr}. The length $L$ of the circumference of
constant $\rho$ in \eqref{metric} is $$L=\rho(2\pi-\triangle
\varphi )$$ where
\begin{equation}\label{deficit angle}
    \triangle\varphi=2\pi\left(4\mu+(1-4\mu)\left[\frac{1}{\rho}\int\limits^{\rho}
    _{\infty}\phi_{1}(\rho')d\rho'-\psi_{1}(\rho)\right]\right)
\end{equation}
is a deficit angle.


\section{Singular magnetic flux}
In this Section we will consider the particular case of a massive
charged scalar field in the background of a singular massless
($\mu=0$) and caring magnetic flux string. Vacuum expectation
value of the induced energy-momentum tensor of a scalar field was
computed in \cite{SG}:
\begin{multline}\label{t0}
\langle T^{t}_{t}\rangle= \langle T^{z}_{z}\rangle=
-\frac{16\sin(F\pi)}{(4\pi)^{3}} \left(\frac
mr\right)^{2}\int\limits_1^\infty
\frac{d\upsilon}{\sqrt{\upsilon^2-1}} \cosh[(2F-1)\,\mathrm
{arccosh}\, \upsilon]\times\\ \times \upsilon^{-3}\left\{
[1+2(1-4\xi)\upsilon^2]K_{2}(2mrv)-2(1-4\xi)mr\upsilon^3
K_{3}(2mr\upsilon)\right\}\,,
\end{multline}
\begin{multline}\label{tr}
\langle T^{r}_{r}\rangle= - \frac{16\sin(F\pi)}{(4\pi)^{3}}
\left(\frac{m}{r} \right)^{2}\int\limits_1^\infty
\frac{d\upsilon}{\sqrt{\upsilon^2-1}} \cosh[(2F-1)\,\mathrm
{arccosh}\, \upsilon]\times\\ \times
\upsilon^{-3}(1-4\xi\upsilon^2) K_{2}(2mr\upsilon)\,,
\end{multline}
\begin{multline}\label{tfi}
\langle T^{\varphi}_{\varphi}\rangle= -
\frac{16\sin(F\pi)}{(4\pi)^{3}}\left(\frac
mr\right)^{2}\int\limits_1^\infty
\frac{d\upsilon}{\sqrt{\upsilon^2-1}} \cosh[(2F-1)\,\mathrm
{arccosh}\, \upsilon]\times\\ \times
\upsilon^{-3}(1-4\xi\upsilon^2)
\left\{K_{2}(2mr\upsilon)-2mr\upsilon K_{3}(2mr\upsilon)\right\},
\end{multline}
where m is the mass of a scalar field, $F$ is the fractional part
($0<F<1$) of the string's magnetic flux $\Phi$ (in the units of
quantum flux $2\pi\hbar/e$) and $\xi$ is the coupling constant of
the scalar field to the scalar curvature of the space-time (for
details see \cite{SG}).

Using  \eqref{tr} and general relations \eqref{phi1},
\eqref{force},
 one can obtain the following expressions for the gravitational force
acting at a point particle of unit mass and for the gravitational
potential:
\begin{multline}\label{gforce}
f(\rho)=\frac{\sin(F\pi)}{\pi^{2}}\cdot
\frac{m^2}{\rho}\int\limits_1^\infty
\frac{d\upsilon}{\sqrt{\upsilon^2-1}} \cosh[(2F-1)\,\mathrm
{arccosh}\, \upsilon]\upsilon^{-3}(1-4\xi\upsilon^2)
K_{2}(2m\rho\upsilon)\,,
\end{multline}
\begin{multline}\label{gpotential}
\phi_1(\rho)=\frac{\sin(F\pi)}{2\pi^{2}}\cdot\frac{m}{\rho}
\int\limits_1^\infty \frac{d\upsilon}{\sqrt{\upsilon^2-1}}
\cosh[(2F-1)\,\mathrm {arccosh}\, \upsilon]
\upsilon^{-4}(1-4\xi\upsilon^2) K_{1}(2m\rho\upsilon).
\end{multline}

The deficit angle \eqref{deficit angle} has the form (see
Appendix):
\begin{multline}\label{deficit}
\triangle\varphi(\rho)=\frac{\sin(F\pi)}{\pi}\cdot\frac{m}{\rho}
\int\limits_1^\infty \frac{d\upsilon}{\sqrt{\upsilon^2-1}}
\cosh[(2F-1)\,\mathrm {arccosh}\, \upsilon] \upsilon^{-4}\times\\
\times\left[(3-2(1-2\xi)\upsilon^2)G(2m\rho
v)+4(1-(1-2\xi)\upsilon^2)K_{1}(2m\rho\upsilon)\right],
\end{multline}
where
\begin{equation}\label{G}
  G(z)=\int \limits^{z}_{\infty} K_0(z)\,dz=\frac{\pi}{2}\left(
  z[K_0(z)L_{-1}(z)+K_{1}(z)L_{0}(z)]-1\right)
\end{equation}
and $L_{\nu}(z)$ is the modified Struve function of order $\nu$
\cite{Prud}.

 Asymptotics of \eqref{gpotential} and \eqref{deficit} at small
 and large distances from the string could be
 easily computed using the asymptotical expressions for
 \eqref{t0}-\eqref{tfi} given in \cite{SG}. For the gravitational
 potential one has
\begin{eqnarray}
 \phi_1(\rho)&\sim&\frac{F(1-F)\gamma(F,\xi)}{12\pi\rho^2}
 \,,\,\,\,\,\,\rho\ll\frac{1}{m}\,,\label{as0}\\
\phi_1(\rho)&\sim&(1-4\xi)\frac{\sin{F\pi}}{8\pi}\cdot
\frac{e^{-2m\rho}}{\rho^2}\,,\,\,\,\,\,\rho\gg\frac{1}{m}\,\label{asinf}
\end{eqnarray}
and for the deficit angle
\begin{eqnarray}
\triangle\varphi(\rho)&\sim&\frac{2F(1-F)\delta(F,\xi)}{3\rho^2}\,,\,\,\,\,\,\rho\ll\frac{1}{m}\,,\label{asdef0}\\
\triangle\varphi(\rho)&\sim&(4\xi-1)\frac{\sin\pi
F}{4}\cdot\frac{e^{-2m\rho}}{\rho^{2}}\,,\,\,\,\,\,\rho\gg\frac{1}{m}\,,\label{asdefinf}
\end{eqnarray}
where we use
\begin{eqnarray}
\gamma(F,\xi)&=&F(1-F)-2(6\xi-1)\,,\label{gamma}\\
\delta(F,\xi)&=&F(1-F)+6\xi-1\,.\label{delta}
\end{eqnarray}

Finally, the general expression \eqref{metric} for the metric in
our case $(\mu=0)$ takes form:
\begin{equation}\label{ourmetric}
    ds^2=\left(1+2\phi_{1}(\rho)\right)[-dt^2+dz^2]+d\rho^2+
    \rho^2\left(1-\frac{\triangle
    \varphi(\rho)}{\pi}\right)d\varphi^2\,,
\end{equation}
where $\phi_{1}(\rho)$ and $\triangle \varphi(\rho)$ are defined
in Eqs.\eqref{gpotential},\eqref{deficit}. Analyzing
Eqs.\eqref{gpotential}, \eqref{deficit} one can conclude that
linear corrections to the  metric has a symmetry $F\leftrightarrow
1-F$ like as components of the energy-momentum tensor
\eqref{t0}-\eqref{tfi}.


\section{Gravitational potential}
Consider the expression for the gravitational potential
\eqref{gpotential}. The function integrated over $\upsilon$ is
product of $K_1(2m\rho\upsilon)$ and $W(\upsilon)$, where
$$W(\upsilon)=\frac{\cosh[(2F-1)\,\mathrm {arccosh}\,
\upsilon]}{\upsilon^{4}\,\sqrt{\upsilon^2-1}}(1-4\xi\upsilon^2)\,.$$
In the region $\upsilon\in[1,\infty)$ $W(\upsilon)$ is negative if
$\xi>1/4$ and once change the sign at
$\upsilon=\frac{1}{2\sqrt{\xi}}$\, if\, $\xi<1/4$. At the same
time $K_1(2m\rho\upsilon)$ is decreasing positive function, which
at $\upsilon\sim\frac{1}{m\rho}$ become less than one and
exponentially goes to zero with increasing $\upsilon$. Analyzing
this product of functions,  one can conclude  that the integral in
\eqref{gpotential} is negative for all values of $\rho$ if
$\xi>1/4$ and may change its sign at some value of $\rho$ if
$\xi<1/4$ but only once.

To clarify  gravitational potential behavior in the region
$\xi<1/4$ one need to analyse sign of the asymptotical expressions
\eqref{as0} and \eqref{asinf}. For $\rho\gg1/m$ one can
immediately get
\begin{equation}\label{condinf}
    \phi_1(\rho)\rightarrow +0\,,\,\,\,\xi<\frac{1}{4}\,\,\,\,\, \rm{and}\,\,\,\,\, \phi_1(\rho)\rightarrow -0\,,\,\,\,\xi
    >\frac{1}{4}\,.
\end{equation}
For knowing sign of $\phi_1(\rho)$ asymptotic at small distances
$(\rho\ll1/m)$ one has to analyse $\gamma(F,\xi)$. Considering it
as a function of $F$ it is easy to see that $\gamma(F,\xi)$ is
negative for all values of $F$ if $\xi>3/16$, positive for all
values of $F$ if $\xi\leq 1/6$ and its sign depends on $F$
otherwise: $\gamma(F,\xi)>0$ if $F\in(F_p\,,1-F_p)$ and
$\gamma(F,\xi)<0$ if $F\in(0,F_p)\cup(1-F_p\,,1)$, where
\begin{equation}\label{fc}
F_p=\frac{1-\sqrt{3(3-16\xi)}}{2}\,.
\end{equation}

Using above,  one can conclude that there are three different
types of gravitational potential behavior:
\begin{description}
    \item[type a.] $\xi\in(1/4,\infty)\,,\quad F\in(0,1)$

     In this case, $\gamma(F,\xi)<0$ at
    all $F$. So at small distances $\phi_1(\rho)$ behaves as
    $-1/\rho^2$ and asymptotically approaches to zero from
    below at large $\rho$. As it can not intersect zero more than once, it can not intersect it at all.
     Analogous we conclude that it has not
    extremes. So, it is monotonic and attractive function at all distances.
     \item[type b.]$\left\{\begin{array}{l}
   \xi\in(3/16,1/4)\,,\quad F\in(0,1)\vspace{0.3em}\\
   \xi\in(1/6,\,3/16]\,,\quad  F\in
   \left(0,F_p\right)\cup\left(1-F_p,1\right)\end{array}\right.$

    In this range of parameters, $\gamma(F,\xi)<0$ and gravitational potential is
    attractive at small distances, but since $\xi<1/4$ it is
    repulsive at large $\rho$. So, with increasing $\rho$, $\phi_1(\rho)$
    increases as $-1/\rho^2$ at $\rho\ll 1/m$, then
    at some $\rho$ intersects zero, reaches its maximum value
    (at $\rho\sim 1/m$ according to numerical computation) and
    decreases to zero from above.
    \item[type c.]$\left\{\begin{array}{l}
   \xi\in(-\infty,1/6]\,,\quad F\in(0,1)\vspace{0.3em}\\
   \xi\in(1/6,\,3/16)\,,\quad F\in\left(F_p,
   1-F_p\right)
   \end{array}\right.$

    In this case,
    $\gamma(F,\xi)>0$ and this means that at small distances
    $\phi_1(\rho)$ behaves as $1/\rho^2$. Since it is the case of
    $\xi<1/4$, then  $\phi_1(\rho)$ asymptotically approaches to zero from
    above at large distances. Analogously to the previous case we conclude that
gravitational potential in this range of
    parameters is repulsive, droningly decreasing function.
 \end{description}

    Our argumentation fails if
    $\xi=1/4$ or
    $F=F_p\,,\,\,F=1-F_p\,$, because in
    this case asymptotical expressions \eqref{as0}, \eqref{asinf} vanishes and we need the next
    terms of expansion.   Frequently considered in the literature
    cases $\xi=0$ (so called minimal coupling) and $\xi=1/6$ (conformal
    coupling) belong to \textbf{type c} that corresponds to repulsion at all distances.

    We plot $\phi_1(\rho)$ (see Fig.1, Fig.2) to see general features
    of the gravitational potential behavior patently.
    Here variable $m\rho$ is along
    $x$-axis and  dimensionless gravitational potential $\phi_1(\rho)/m^2$ is along
    $y$-axis. In the regions where type of the gravitational potential behavior
    does
    not depend on the magnetic flux $F$, the $\xi-$dependence of the gravitational
    potential is presented in Fig.1. In the region where type of
    gravitational potential behavior is sensitive to the magnetic
    flux $\left(\xi\in(1/6,\,3/16)\right)$, we
    illustrate the gravitational potential as $F$ function in Fig.2.

    The maximum amplitude of the local
    gravitational potential is at half-integer value of flux
    $(F=1/2)$. In the $F$-sensitive alternating-sign part of the region
    ($1/6<\xi<3/16$,
    $F\in\left(0,F_p\right)\cup\left(1-F_p\,,1\right)$),
      effect is increasing under going from zero flux to
     the border point $F=F_p$ (and from $F=1$ to
     the border point $F=1-F_p$).

Condition \eqref{cond of validity} of validity of our linear
approximation is violated near the string. Using asymptotical
expression \eqref{as0} for the gravitational potential at small
distances and recovering the dimension, one can rewrite the
condition of validity in the form
\begin{equation}\label{condition of validity}
     \frac{l_p^2}{\rho^2}\ll 1\,,
\end{equation}
where $l_p\sim 10^{-33}cm$ is Plank length. This is what we
expected since, at Plank scales, the semiclassical approach
\eqref{sma} is violated, and we can not consider the gravitational
field as a background of quantum processes.

Finishing this Section it will be useful to note that since
gravitational force \eqref{gforce} is a derivative of
\eqref{gpotential}, the gravitational force behavior is similar to
the behavior of gravitational potential.


 \section{Deficit angle}
The middle distance behavior of the deficit angle \eqref{deficit}
is not so clear as for the gravitational potential
\eqref{gpotential}. So one can differentiate the  deficit angle
behavior at three types depending on the asymptotical behavior.
Using \eqref{asdef0}-\eqref{asdefinf}we obtain:
\begin{description}

\item[type 1]\, $\xi\in(1/4,\infty)\,,\quad F\in(0,1)$

Deficit angle is positive at small and large distances. At small
distances it behaves as $1/\rho^2$ and exponentially decreases at
$\rho\gg 1/m$.

\item[type 2] $\left\{\begin{array}{l}
   \xi\in[1/6,1/4)\,,\quad F\in(0,1)\vspace{0.3em}\\
   \xi\in(1/8,\,1/6)\,,\quad  F\in(F_d,1-F_d)\end{array}\right.$

At distances $\rho\ll 1/m$ the deficit angle is positive  and
behaves as $1/\rho^2$, but at large distances it change its sign
(at least once) and approaches to zero from below.

\item[type 3] $\left\{\begin{array}{l}
   \xi\in(-\infty,1/8)\,,\quad F\in(0,1)\vspace{0.3em}\\
   \xi\in[1/8,\,1/6)\,,\quad   F\in
   \left(0,F_d\right)\cup\left(1-F_d,1\right)\end{array}\right.$

Deficit angle is negative at small and large distances.

\end{description}

Here we used notation
\begin{equation}\label{Fd}
  F_d=\frac{1-\sqrt{3(8\xi-1)}}{2}\,.
\end{equation}

    We plot $\triangle\varphi(\rho)$ (see Fig.3, Fig.4) to see the general features of
    the deficit angle behavior evidently.
    Here variable $m\rho$ is along
    $x$-axis and the dimensionless deficit angle $\triangle\varphi(\rho)/m^2$ is along
    $y$-axis. In the regions where type of the deficit angle behavior
    does
    not depend on the magnetic flux $F$, the $\xi-$dependence of the deficit angle
     is presented in Fig.3. In the region where type of
    the deficit angle behavior is sensitive to the magnetic
    flux $\left(\xi\in(1/8,\,1/6)\right)$, we
    illustrate  the deficit angle as $F$ function in Fig.4.

    The  maximum amplitude of local
    deficit angle is at the half-integer value of the flux
    $(F=1/2)$ except the $F$-sensitive
    part of region ($1/8<\xi<1/6$) where  if $F\in(F_d\,,1-F_d)$
    the effect is minimal at $F=1/2$ and take its maximum peak
    at the border points $F=F_d$; for the case $F\in
   \left(0,F_d\right)\cup\left(1-F_d,1\right)$ the lines
   corresponding different values of $F$ goes
   to negative infinity near the string and
   intersects under moving away from one.

Condition of validity of our result for the deficit angle
coincides with \eqref{condition of validity} and, hence, do not
lead to the new restrictions.


\section{Concluding remarks}
In the semiclassical approach, we computed the gravitational
effect caused by the vacuum polarization of the massive charged
scalar field in the background of a singular massless carrying
magnetic flux cosmic string. Corrections \eqref{gpotential},
\eqref{deficit}  to the metric components \eqref{ourmetric} depend
periodically on the cosmic string flux $(\Phi)$ (i.e. depends on
only its fractional value $F$), has a symmetry $F\leftrightarrow
1-F$ and vanish at its integer value $(\Phi=n)$.

It turned out, that behavior of the gravitational potential can be
divided at 3 types depending on $F$ and coupling constant $\xi$.
This three types are: attractive behavior over all distances,
repulsive behavior\footnote{There is no wonder in a repulsive
behavior of the gravitation potential  at some values of
parameters because of violation of the strong and week energy
conditions for components of the induced energy-momentum tensor of
the massive charged scalar field \eqref{t0}-\eqref{tfi} (see
\cite{SG}).} over all distances and alternating-sign behavior
(gravitational potential is negative near the string but change
its sign and becomes positive under moving away from it). The
areas of $F$ and $\xi$ parameters for different types are pointed
out in Section 4. Gravitational potential was found to be
repulsive in the commonly considered cases of $\xi=0$ and
$\xi=1/6$.

To see general features of gravitational potential behavior
patently we plot $\phi_1(\rho)$ (see Fig.1, Fig.2). Here variable
$m\rho$ is along $x$-axis and dimensionless gravitational
potential $\phi_1(\rho)/m^2$ is along $y$-axis. In the regions
where type of the gravitational potential behavior does not depend
on the magnetic flux $F$, the $\xi-$dependence of the
gravitational potential is presented in Fig.1. In the region where
type of the gravitational potential behavior is sensitive to the
magnetic flux $\left(\xi\in(1/6,\,3/16]\right)$, we illustrate the
gravitational potential as $F$ function in Fig.2.

The behavior of the deficit angle can be also divided in 3 types.
The areas of $F$ and $\xi$ parameters for different types are
pointed out in Section 5. To see general features of the deficit
angle behavior patently we plot $\triangle\varphi(\rho)$ (see
Fig.3, Fig.4). Here variable $m\rho$ is again along $x$-axis and
dimensionless deficit angle $\triangle\varphi(\rho)/m^2$ is along
$y$-axis. In the regions where type of the deficit angle behavior
does not depend on the magnetic flux $F$, the $\xi-$dependence of
the deficit angle is presented in Fig.3.  In the region where type
of the deficit angle behavior is sensitive to magnetic flux
$\left(\xi\in[1/8,\,1/6)\right)$, we plot the deficit angle as $F$
function in Fig.4.

It is interesting to note that regions of the different types of
the gravitational potential and deficit angle behavior strictly
speaking are different. From the classical point of view we could
expect that for the attractive-type potentials deficit angle is
positive and for repulsive-type potentials one is negative
(positive deficit angle lead to the bending of light as like it
attracts to the string and vice versa). This is a fact for the
regions $\xi>1/4$ (attractive-type potential and positive deficit
angle) and  $\xi<1/8$ (repulsive-type potential and negative
deficit angle). But at the region $\xi\in(1/8,1/4)$ our classical
reasons fail. For example, if $\xi=1/6$ (conformal coupling), the
gravitational potential of a string is repulsive at all distances,
while the deficit angle is alternating-sign.

Near a string the potential and deficit angle behaves as like the
scalar field is massless and we recover the result of \cite{G}
(for the zero linear mass density of string). But in comparing
with \cite{G}\footnote{In \cite{G}, the gravitational potential
and the deficit angle can be only positive or negative over all
distances from string.} (see asymptotic expressions \eqref{as0},
\eqref{asdef0}), the massiveness of field gives an essentially new
type of behavior that allow gravitational potential and deficit
angle change its sign under moving away from the string. One
another difference is that the massless scalar field produces a
long-range power decreasing potential \eqref{as0} and deficit
angle \eqref{asdef0}, while in our case it are short-range
exponentially decreasing functions \eqref{asinf} and
\eqref{asdefinf}.

Condition of validity of semiclassical approximation
\eqref{condition of validity} is violated near the string at Plank
length. It should be noted that we considered simplified
analytically solved case of massless $(\mu=0)$ string with zero
radius. For realistic  string which radius is of the order of the
Compton wavelength of the Higgs bosons involved in the phase
transitions, the condition \eqref{condition of validity} is
satisfied everywhere outside the string.

As was pointed in \cite{G}, the contribution to the gravitational
effect coming from the Aharonov-Bohm interaction dominates over
one coming from the nonzero linear mass density $\mu$ of the
string. So we can expect that our results will be not changed
significantly with taking in to account $\mu$.

\section*{Acknowledgements}
We are grateful to Profs. Yu.A. Sitenko and Yu.V. Shtanov for
invaluable help in preparing paper, useful discussions and
critical reading. A.V. acknowledges support from BITP Educational
Center and personally V. Shadura.


{ \setcounter{equation}{0}
\renewcommand{\theequation}{A.\arabic{equation}}
\section*{Appendix}

The deficit angle is given by \eqref{deficit angle} . Consider
separately two corresponding terms (in our case $\mu=0$).

Using recurrent relation
\begin{equation}\label{requrent}
\frac{K_1(z)}{z}=-K_0(z)-\partial_z K_1(z)
\end{equation}
and Eq.\eqref{gpotential} for $\phi_1(\rho)$ one can easily obtain
\begin{multline}\label{IntPhi1}
\frac{1}{\rho}
\int\limits^{\rho}_{\infty}\phi_1(\rho')d\rho'=-\frac{\sin(F\pi)}{2\pi^{2}}\cdot
   \frac{m}{\rho}\int\limits_1^\infty \frac{d\upsilon}{\sqrt{\upsilon^2-1}}
\cosh[(2F-1)\,\mathrm {arccosh}\, \upsilon]
\upsilon^{-4}(1-4\xi\upsilon^2)\times\\ \times[G(2m\rho\upsilon)+
K_{1}(2m\rho\upsilon)],
\end{multline}
where $G(z)$ is defined in \eqref{G}.

To compute  $\psi_1(\rho)$ one need expression under integral
operation in \eqref{psi1}:
\begin{multline}\label{middle2}
2\langle T^{t}_{t}\rangle-\langle T^{\varphi}_{\varphi}\rangle= -
\frac{\sin(F\pi)}{4\pi^{3}}\left(\frac
m\rho\right)^{2}\int\limits_1^\infty
\frac{d\upsilon}{\sqrt{\upsilon^2-1}} \cosh[(2F-1)\,\mathrm
{arccosh}\, \upsilon]\times\\ \times
\upsilon^{-3}\left\{[1+4(1-3\xi)\upsilon^2]K_{2}(2m\rho\upsilon)+[1-2(1-2\xi)\upsilon^2]\cdot2m\rho\upsilon\cdot
K_{3}(2m\rho\upsilon)\right\}.
\end{multline}

Using relation
\begin{multline}\label{Helppsi1}
\int[b_1K_2(z)+b_2zK_3(z)]\,dz=\int[(b_1+2b_2)K_2(z)-b_2z\partial_zK_2(z)]\,dz=\\
=-\{b_2zK_2(z)+(b1+3b_2)[G(z)+2K_1(z)]\},
\end{multline}
 and Eq.\eqref{middle2} one can get:
\begin{multline}\label{intpsi1}
\int\limits^{\rho'}_{\infty}d\rho
    ''\cdot \rho''^{2}[2\langle T^{t}_{t}(\rho'')\rangle-\langle
    T^{\varphi}_{\varphi}(\rho'')\rangle]=
    \frac{\sin(F\pi)}{8\pi^{3}}\,
m\int\limits_1^\infty \frac{d\upsilon}{\sqrt{\upsilon^2-1}}
\cosh[(2F-1)\,\mathrm {arccosh}\, \upsilon]\times\\ \times
\upsilon^{-4}\left\{[1-2(1-2\xi)\upsilon^2]\cdot2m\rho'\upsilon\cdot
K_{2}(2m\rho'\upsilon)+2(2-\upsilon^2)[G(2m\rho'\upsilon)+2K_{1}(2m\rho'\upsilon)]\right\}.
\end{multline}

Using expression
\begin{multline}\label{helppsi2}
\int\frac{dz}{z^2}[f_1zK_2(z)+f_2(G(z)+2K_1(z))]= \int
dz\left[\frac{K_2(z)}{z}(f_1+f_2)+f_2\left(\frac{G(z)}{z^2}-\frac{K_0(z)
}{z}\right)\right]=\\
\\=\int dz\left[\frac{K_2(z)}{z}(f_1+f_2)-f_2\partial_z\left(\frac{G(z)}{z}\right)\right]=
-\frac{1}{z}\left[(f_1+f_2)K_1(z)+f_2G(z)\right]
\end{multline}
and Eqs.\eqref{intpsi1}, \eqref{psi1} one can easily obtain:
\begin{multline}\label{psi1finish}
\psi_{1}(r)= -\frac{\sin(F\pi)}{2\pi^{2}}\,
\frac{m}{\rho}\int\limits_1^\infty
\frac{d\upsilon}{\sqrt{\upsilon^2-1}} \cosh[(2F-1)\,\mathrm
{arccosh}\, \upsilon]\times\\ \times
\upsilon^{-4}\left\{(5-4(1-\xi)\upsilon^2)
K_{1}(2m\rho\upsilon)+2(2-\upsilon^2)G(2m\rho\upsilon)\right\},
\end{multline}

After substitution \eqref{IntPhi1}, \eqref{psi1finish} into
\eqref{psi1} one gets \eqref{deficit}.

        \begin{figure}
        \begin{tabular}{c}
        \includegraphics[width=140mm]{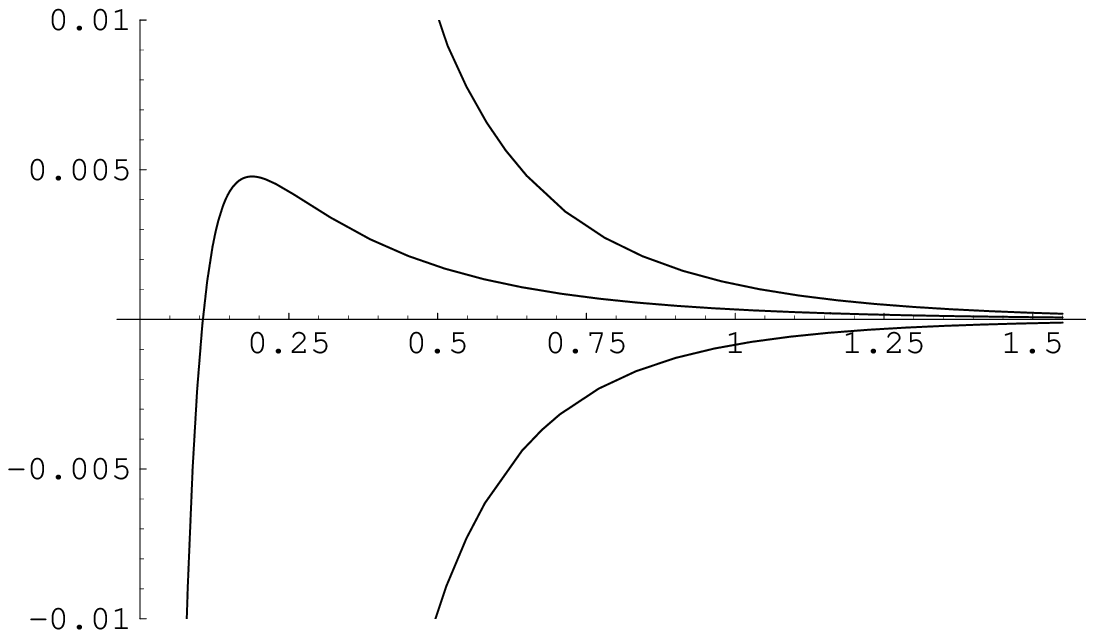}
        \end{tabular}
        \caption{Gravitational potential in $F$-independent regions for  $F=1/2$ and
        $\xi=0.26$ (type \textbf{a}), $0.19$ (type \textbf{b}), $0.14$ (type \textbf{c})
        correspondingly from down to up.}
        \begin{tabular}{c}
        \includegraphics[width=140mm]{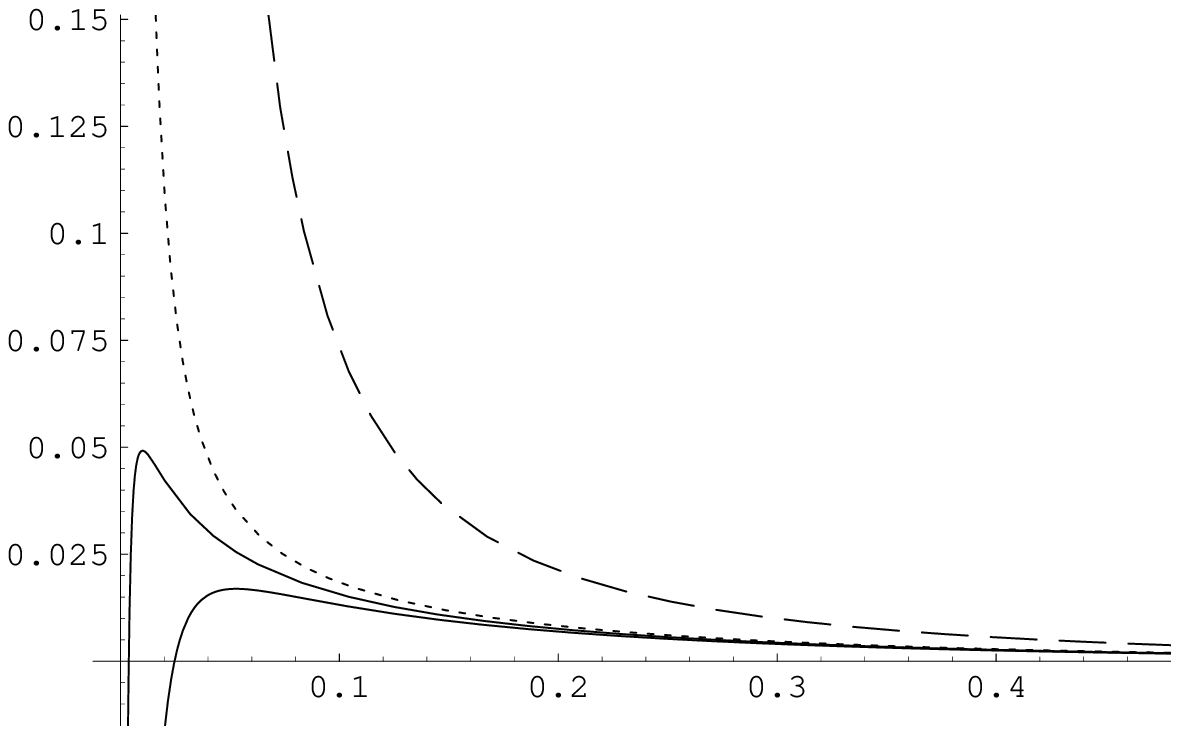}
        \end{tabular}
        \caption{Gravitational potential at the $F-$sensitive
        region
        $\left( \xi\in(\frac{1}{6},\frac{3}{16})\right)$ for the cases of
        $\xi=0.18$, $F_p=0.2$. Solid lines correspond
        $F$ from the region
        $F\in\left(0,F_p\right)\cup\left(1-F_p,1\right)$:
        $F=0.19,\,\,0.1995$
        accordingly from down to up.
        For the region $F\in\left(F_p,1-F_p\right)$ dotted line corresponds to $F=0.21$, dashed line to $F=1/2$.}
        \end{figure}

        \begin{figure}
        \begin{tabular}{c}
        \includegraphics[width=140mm]{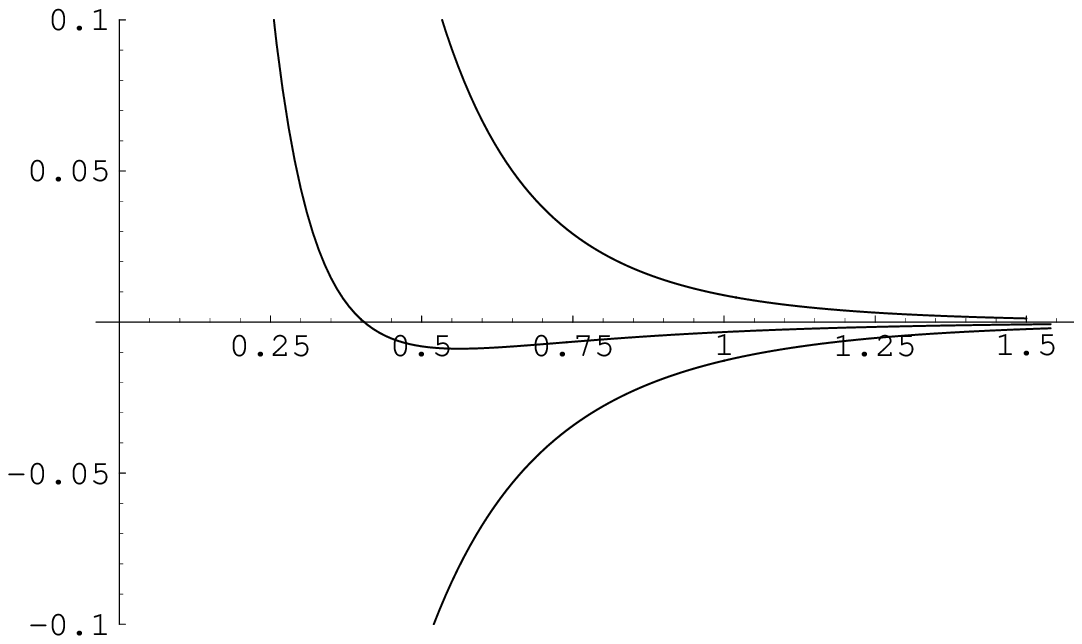}
        \end{tabular}
        \caption{Deficit angle in $F$-independent regions for  $F=1/2$ and
        $\xi=0.26$ (type \textbf{1}), $0.17$ (type \textbf{2}), $0.1$ (type \textbf{3})
        correspondingly from up to down.}
        \begin{tabular}{c}
        \includegraphics[width=140mm]{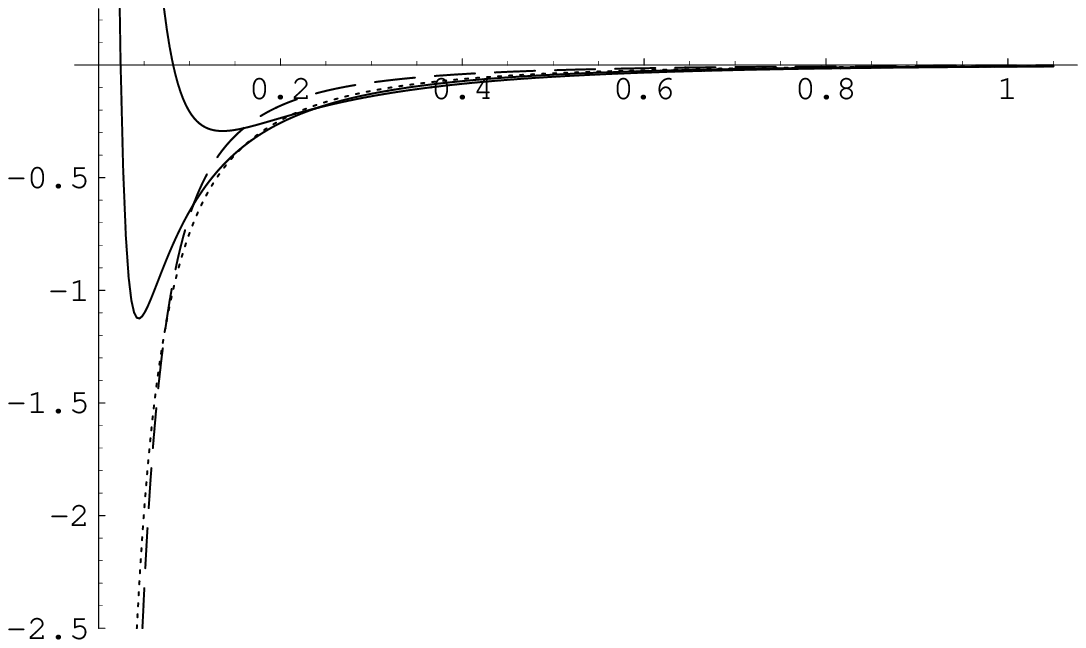}
        \end{tabular}
        \caption{Deficit angle at the $F-$sensitive
        region
        $\left( \xi\in(\frac{1}{8},\frac{1}{6})\right)$ for the cases of
        $\xi=0.14$, $F_d=0.2$. Solid lines correspond
        $F$ from the region  $F\in(F^d_{c},1-F_d)$: $F=0.5,\,\,0.25$
        correspondingly from up to down.
        For the region $F\in \left(0,F_d\right)\cup\left(1-F_d,1\right)$ dotted line corresponds to $F=0.19$,
         dashed line to $F=0.1$.}
        \end{figure}
  \end{document}